\documentclass[prb,amsmath,amssymb,twocolumn,showpacs]{revtex4-1}
\usepackage{graphicx}
\usepackage{tabularx}
\usepackage{color}
\usepackage{amsmath}
\usepackage{mathtools}
\usepackage{comment}
\newcommand{\be}{\begin{equation}}
\newcommand{\ee}{\end{equation}}
\newcommand{\bea}{\begin{eqnarray}}
\newcommand{\eea}{\end{eqnarray}}
\newcommand{\ket}[1]{\left| #1 \right>} 
\usepackage{dcolumn}
\usepackage{hyperref}
\usepackage{bm}
\usepackage{epsf}
\usepackage{subfigure}
\usepackage{epstopdf}%
\usepackage{amsfonts}
\usepackage{braket}
\usepackage{cancel}

\begin{document}

\title{Current fluctuations in  quantum absorption refrigerators}
\author{Dvira Segal}
\affiliation{
Department of Chemistry and Centre for Quantum Information and Quantum Control,
University of Toronto, 80 Saint George St., Toronto, Ontario, Canada M5S 3H6
}
\date{\today}

\begin{abstract}
Absorption refrigerators transfer thermal energy from a cold bath to a hot bath
without input power by utilizing heat from an additional ``work" reservoir.
Particularly interesting is a three-level design for a
quantum absorption refrigerator, which can be optimized to reach the maximal (Carnot) cooling efficiency.
Previous studies of three-level chillers focused on the behavior of the averaged cooling current.
Here, we go beyond that 
and study the full counting statistics of heat exchange in a three-level chiller model.
We explain how to obtain the complete cumulant generating function of the refrigerator in steady state, then
derive a partial cumulant generating function, which yields 
closed-form expressions for both the averaged cooling current and its noise.
Our analytical results and simulations are beneficial
for the design of nanoscale engines and cooling systems far from equilibrium, with their performance
optimized according to different criteria, efficiency, power, fluctuations and dissipation.
\end{abstract}

\maketitle

\section{Introduction}
\label{sec-intro}

Autonomous absorption refrigerators transfer thermal energy from a cold ($c$)
bath to a hot ($h$) bath by utilizing heat from an ultra-hot heat bath,
termed as a work ($w$) reservoir.
Absorption refrigerators were realized in the 19th century~\cite{AR19},
and their analysis was central to the
development of the theory of finite-time irreversible thermodynamics.
Theoretical studies of quantum, nanoscale analogues of absorption refrigerators
aim to establish the theory of macroscopic thermodynamics from quantum principles
\cite{review1, kos13,reviewARPC14,Goold,kos18}.
Recent experiments demonstrated classical nanoscale engines,  e.g.
a single atom heat engine \cite{singleatom} and a nanomechanical heat engine utilizing squeezed thermal 
reservoirs \cite{squeezeE}. 
More recently, a quantum absorption refrigerator was realized using three trapped ions 
\cite{QARE}.
%

\begin{figure}[htbp]
\includegraphics[width=7cm]{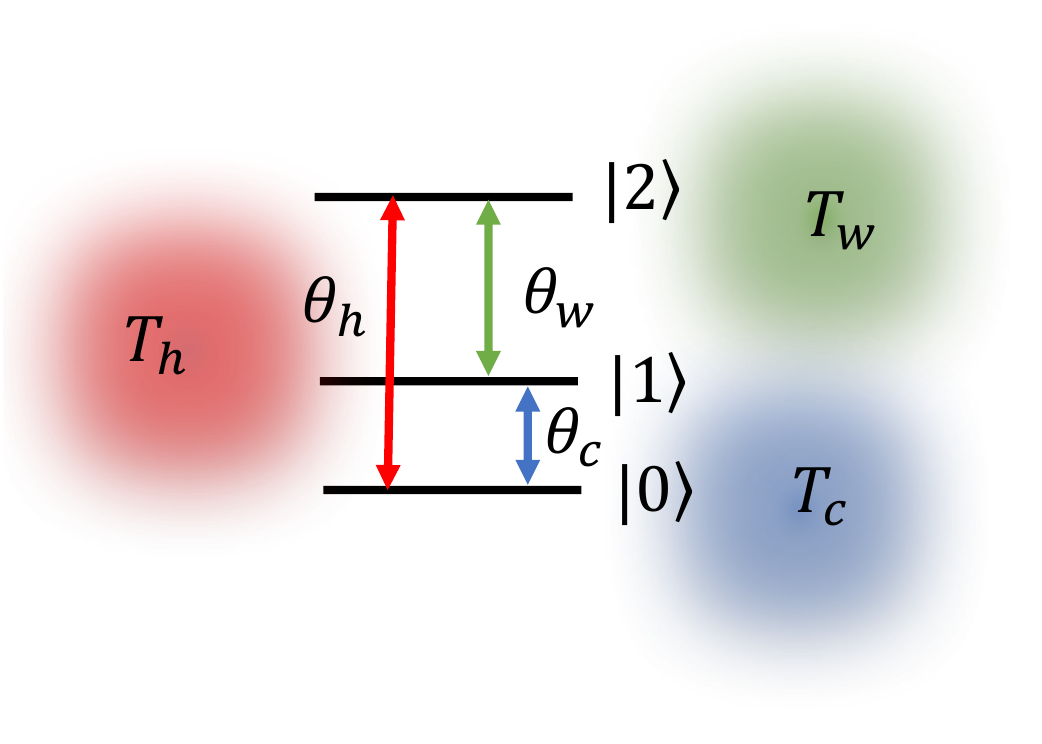}
\vspace{-7mm}
\caption{A three-level quantum absorption refrigerator. 
It includes a quantum system (working fluid) with
each transition coupled to an independent thermal bath, hot ($h$), cold ($c$)  and work ($w$).
}
\label{FigQAR}
\end{figure}

An illustrious design of an autonomous quantum absorption refrigerator (QAR)
consists of a three-level system as the working fluid and
three independent thermal reservoirs \cite{reviewARPC14,joseSR}.
Each transition between a pair of levels is coupled to one of the three
heat baths, $c$, $h$ and $w$, where  $T_w>T_h>T_c$.
For a schematic representation, 
see Fig.~\ref{FigQAR}.
In the steady state limit, the work bath provides energy to the system
promoting the extraction of energy
from the cold bath, to be dumped into the hot reservoir. The opposite
heating process, from the hot bath to the cold one, can be controlled and played down
by manipulating the frequencies of the system.

The three-level QAR and its variants were discussed in details in several recent studies,
see e.g. Refs. \cite{reviewARPC14,joseSR, Levy12,Linden11,plenio,PopescuPRL,Popescu12,Alonso14,jose15,AlonsoNJP}.
It is designed to perform optimally under the weak coupling approximation,
when each bath individually couples to a different transition within the system.
Quantum coherences are expected
to negatively impact the cooling performance of multilevel QARs by introducing
internal dissipation and leakage processes \cite{jose15}.
Off-resonant effects and bath-cooperative interactions may contribute to energy exchange
once the system strongly interacts with the thermal baths.
Such nonlinear effects are difficult to control and can lead to a rather poor performance.
In a recent study we had analyzed a {\it qubit} QAR by making the system to strongly
interact with three shaped reservoirs \cite{Anqi}. Carnot efficiency was then achieved in a
singular design---with the baths engineered to consist only specific frequency components.
The three-level QAR and similar continuous (non-reciprocating) machines 
can be realized in different physical systems including coupled atoms and ions, multi-site electronic junctions
and photovoltaic cells, as described e.g. in Refs. \cite{plenio,NitzanE1,KilgourD}.


\vspace{-2mm}

Beyond the analysis of the averaged cooling current, a full counting statistics (FCS)
formalism provides the fundamental description of an out-of-equilibrium
quantum system \cite{esposito-review,hanggi-review,bijay-wang-review}.
Such an approach hands over the cumulant generating function (CGF) for heat exchange,
fully characterizing energy transport in steady state.
The goal of the present paper is to employ a full counting statistics formalism and
study the cooling performance of a three-level QAR.

The paper includes three central contributions:
(i) We describe the calculation of the FCS of heat exchange in the three-level QAR.
(ii) We introduce a practical method for reaching closed-form expressions for the first two
cumulants (cooling current and its noise) through the derivation of a {\it partial} CGF.
The method is general and can be readily applied to other multilevel models
beyond the weak coupling approximation.
(iii) We simulate the behavior of the cooling current, its noise and the accuracy-dissipation trade off. 
Altogether, these measures are important for the optimization of heat machines \cite{vanden16,TUR1,TUR2,TUR3a,TUR3b,Geissler}.

The work is organized as follows. We introduce our model in Sec. \ref{Smodel}.
In Sec.~\ref{SFCS} we present the FCS approach and
derive expressions for the cooling current and its noise.
We illustrate our results with numerical simulations in Sec. \ref{Ssimul} and
summarize in Sec.~\ref{Ssum}.


\section{Model}
\label{Smodel}

The paradigmatic three-level QAR is described by the Hamiltonian  $\hat H=\hat H_s+\hat H_b + \hat H_{sb}$ with
a three-level system $\hat H_s=\sum_{0,1,2}E_{j}|j\rangle \langle j|$
and three reservoirs $\hat H_b=\hat H_c +\hat H_w + \hat H_h$.
The baths include collections of independent harmonic oscillators,
$\hat H_{\nu}=\sum_{k}\omega_{\nu,k}a_{\nu,k}^{\dagger}a_{\nu,k}$,
with
$a_{\nu,k}^{\dagger}$ ($a_{\nu,k}$) the creation (annihilation) 
operator of mode $k$ with frequency $\omega_{\nu,k}$
in the $\nu=h,c,w$ bath. 
Each thermal bath is coupled to a specific transition within the system,
\bea
\hat H_{sb}&=& \hat B_c\left(|0\rangle\langle 1| +|1\rangle\langle 0| \right)
  + \hat B_{w}\left(|1\rangle\langle 2| +|2\rangle\langle 1| \right)
\nonumber\\
&+&\hat B_{h}\left(|0\rangle\langle 2| +|2\rangle\langle 0| \right).
\eea
The bath operators are 
$\hat B_{\nu}=\sum_{k}g_{\nu,k}\left(a_{\nu,k}^{\dagger}+a_{\nu,k}\right)$,
but the analysis can be readily generalized beyond that.
In what follows we use the notation $\theta_c\equiv E_1-E_0$,
$\theta_w\equiv E_2-E_1$ and $\theta_h\equiv\theta_c+\theta_w$.

The dynamics of the reduced density matrix can be organized
as a Markovian quantum master equation for the levels' populations 
under the following assumptions \cite{Breuer}:
(i) weak system-bath coupling, (ii) Markovian reservoirs, and (iii) 
decoupled coherences-population dynamics (secular approximation).
This standard scheme results in the following kinetic-like equations
\bea
\dot p_{0}&=&-\left(k_{0\to 1}+k_{0\to 2}\right) p_0(t) +  k_{1\to 0} p_1(t) + k_{2\to 0} p_2(t),
\nonumber\\
\dot p_{1}&=&-\left(k_{1\to 0}+k_{1\to 2}\right) p_1(t) +  k_{0\to 1} p_0(t) + k_{2\to 1} p_2(t),
\nonumber\\
\dot p_{2}&=&-\left(k_{2\to 0}+k_{2\to 1}\right) p_2(t) +  k_{0\to 2} p_0(t) + k_{1\to 2} p_1(t),
\nonumber\\
\label{eq:pop}
\eea
with rate constants
\bea
k_{0\to 1} &=& \Gamma_c(\theta_{c}) n_c(\theta_c),\,\,\,
k_{0\to 2} = \Gamma_h(\theta_h) n_h(\theta_h),
\nonumber\\
k_{1\to 2} &=& \Gamma_w(\theta_w) n_w(\theta_w).
\eea
The detailed balance relation dictates the rate constants of reversed processes, e.g.,
$k_{1\to 0} =  \Gamma_c(\theta_{c}) \left[n_c(\theta_c)+1\right]$.
The system-bath coupling constants (hybridization) are
$\Gamma_{\nu}(\theta_{\nu})=2\pi \sum_{k}|g_{\nu,k}|^2\delta(\omega_{\nu,k}-\theta_{\nu})$,
and we used an ohmic function to model them,
$\Gamma_{\nu}(\theta_{\nu})=\theta_{\nu}$.
The  Bose-Einstein occupation factors are $n_{\nu}(\theta_{\nu})=[e^{\beta_{\nu}\theta_{\nu}}-1]^{-1}$,
given in terms of the inverse temperature $\beta_{\nu}=1/(k_BT_{\nu})$. 
In the weak-coupling approximation employed here only resonance processes are allowed.
For brevity, we omit below the reference to frequency in
 $\Gamma_{\nu}(\theta_{\nu})$ and $n_{\nu}(\theta_{\nu})$. 


Our equations of motion can be written in terms of the dissipators $L_{\nu}$ as
 $|\dot p\rangle= (L_c+L_h+L_w)|p(t)\rangle$, with $| p\rangle$  a vector of population and $\sum_jp_j(t)=1$.
This equation can be solved in the long time limit, yielding the steady state population, $p_j^{ss}$.
The rate of change of the system's energy is given by
 $d\langle \hat H_s\rangle/dt$. 
From here, we can immediately write down a closed-form expression for the averaged energy current at
the $\nu$ contact, namely $\langle J_{\nu}\rangle= \sum_j (L_{\nu}|p^{ss}\rangle)_j E_j$. 
This procedure was employed in previous studies, see e.g. \cite{reviewARPC14,joseSR}.
In the next section we generalize this description by using a full counting statistics analysis, and 
explicitly calculate the first two cumulants.


\section{Full Counting Statistics}
\label{SFCS}

\subsection{The complete CGF}

The cumulant generating function of the model described in Sec. \ref{Smodel}
can be derived by following a rigorous procedure \cite{esposito-review,hava}.
Here, for simplicity, we employ a classical, intuitive derivation of the FCS; it complies with the rigorous method 
under the weak-coupling, Markovian and secular approximations \cite{hava}.
We begin by defining $\mathcal{P}_t(j,n\theta_c,m\theta_w)$ 
as the probability that by the time (long time) $t$,
a total energy $n\theta_c$ has been absorbed by the system 
from the cold bath, $m\theta_w$ energy has been transferred from the work bath to the system,
and the system populates the state $j=0,1,2$. Since the dynamics is Markovian and energy is conserved,
we do not count energy at the third bath.
Note that $n,m$ are integers
since  under the weak coupling (resonant transmission)
approximation, heat is transferred into and out of the system in discrete quanta $\theta_{c,w,h}$.
We can readily write down an equation of motion for $\mathcal{P}_t(j,n\theta_c,m\theta_w)$ 
\cite{esposito-review,Renjie,yelenaPRB,yelenaCGF},
\begin{widetext}
\bea 
\dot{ \mathcal P}_t(0,n\theta_c,m\theta_w)
& = &
- \mathcal P_t(0,n\theta_c,m\theta_w) (k_{0\rightarrow 1} +  k_{0\rightarrow 2} ) +
\mathcal P_t(1,(n+1)\theta_c,m\theta_w) k_{1\rightarrow 0} +
\mathcal P_t(2,n\theta_c,m\theta_w) k_{2\rightarrow 0},
\nonumber\\
\dot{ \mathcal P}_t(1,n\theta_c,m\theta_w)
& = & - \mathcal P_t(1,n\theta_c,m\theta_w)
(k_{1\rightarrow 0} +  k_{1\rightarrow 2} ) +
\mathcal P_t(0,(n-1)\theta_c) k_{0\rightarrow 1} +
\mathcal P_t(2,n\theta_c,(m+1)\theta_w)k_{2\rightarrow 1},
\nonumber\\
\dot{ \mathcal P}_t(2,n\theta_c,m\theta_w)
& = & - \mathcal P_t(2,n\theta_c,m\theta_w)
(k_{2\rightarrow 0} +  k_{2\rightarrow 1} ) +
\mathcal P_t(1,n\theta_c,(m-1)\theta_w) k_{1\rightarrow 2} +  \mathcal P_t(0,n\theta_c,m\theta_w)
k_{0\rightarrow 2}.
\label{eq:Pwweak}
\eea
\end{widetext}

\begin{figure*}[htbp]
\includegraphics[width=16cm]{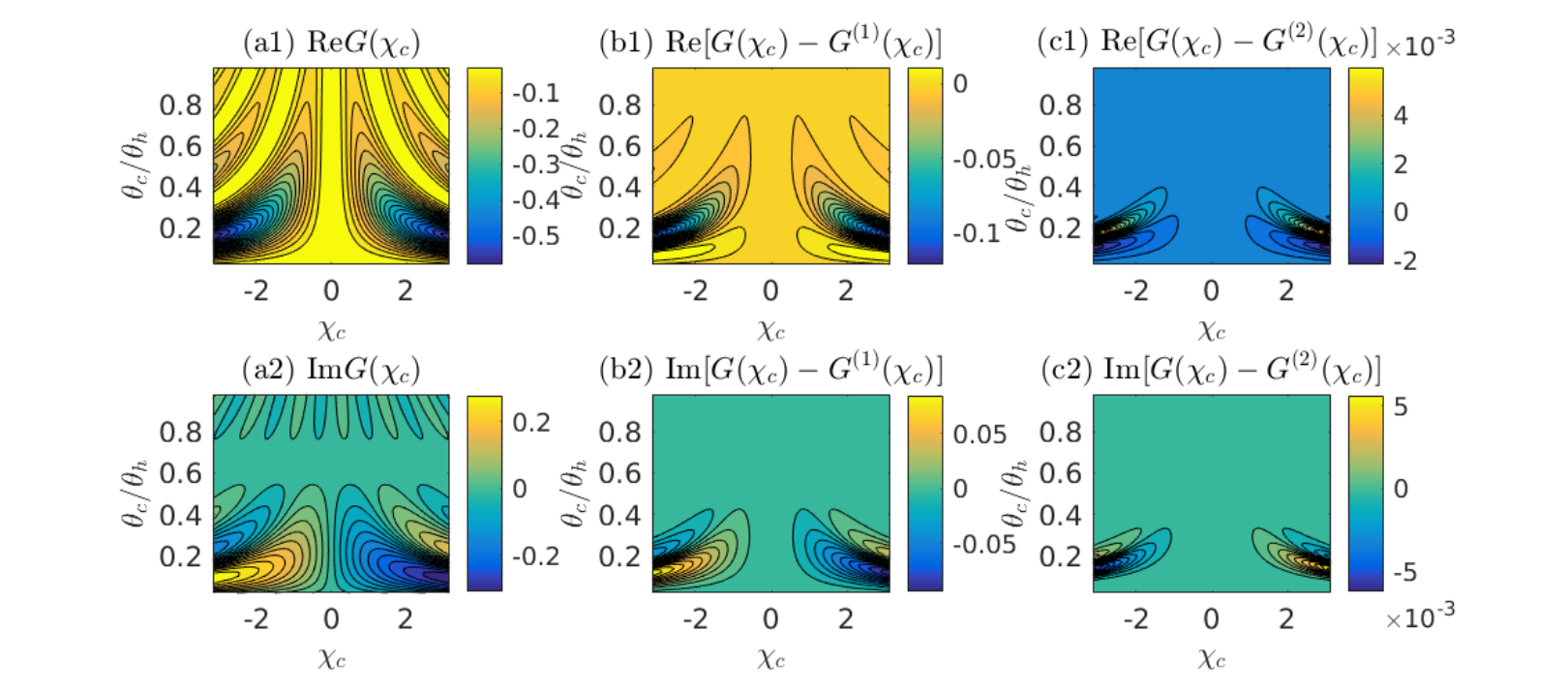} 
\caption{Real (Re) and imaginary (Im) parts of the cumulant generating function $G(\chi_c,\chi_w=0)$.
(a) Complete CGF computed from a numerical solution of the eigenvalue problem, Eqs. (\ref{eq:ZW})-(\ref{eq:lam1}).
(b) Deviation of the first order CGF $G^{(1)}(\chi)$ of Eq. (\ref{eq:G1})
from the complete CGF.
(c) Deviation of the second order CGF $G^{(2)}(\chi)$ of Eq. (\ref{eq:G2})
from the complete CGF.
We used $\beta_c=1$, $\beta_w=0.2$,  $\beta_h=0.7$, $\theta_h=6$, and we varied $\theta_c$ 
up to $\theta_h$.
}
\label{FigCGF}
\end{figure*}

We Fourier-transform these equations by introducing the the so-called counting fields 
$\chi=(\chi_c,\chi_w)$, and obtain the characteristic function  
\bea
&&\ket{Z(\chi,t)} \equiv
\nonumber\\
&&\begin{pmatrix}
\sum_{n,m=-\infty}^{\infty} \mathcal P_t(0,n\theta_c,m\theta_w)e^{in\theta_c\chi_c}  e^{im\theta_w\chi_w}\
\\ \sum_{n,m=-\infty}^{\infty} \mathcal
P_t(1,n\theta_c,m\theta_w)e^{in\theta_c\chi_c}  e^{im\theta_w\chi_w}\
\\ \sum_{n,m=-\infty}^{\infty} \mathcal
P_t(2,n\theta_c,m\theta_w)e^{in\theta_c\chi_c} e^{im\theta_w\chi_w}\
\end{pmatrix}
\nonumber\\
\label{eq:zW} \eea
It satisfies a first order differential equation, 
\bea \frac{ d| Z(\chi, t)\rangle}{dt}=
\hat W(\chi)|Z(\chi, t)\rangle
\label{eq:ZW}
\eea
with the rate matrix 
\bea \hat W=
\begin{pmatrix}
-k_{0\rightarrow 1}-k_{0\rightarrow 2}  &  k_{1\rightarrow 0} e^{-i\chi_c\theta_c}   & k_{2\rightarrow 0}  \\
k_{0\rightarrow 1} e^{i\chi_c \theta_c}  &   -k_{1\rightarrow 0} - k_{1\rightarrow 2 } & k_{2\to 1} e^{-i\chi_w\theta_w}\\
k_{0\rightarrow 2}   &   k_{1\rightarrow 2} e^{i\chi_w\theta_w} &  -k_{2\rightarrow 1 } -k_{2\to 0}\\
\end{pmatrix}
\nonumber\\
\label{eq:M}
\eea
The characteristic polynomial of  $\hat W$ is written schematically as 
\bea
&&\lambda^3-a_1\lambda^2 + a_2\lambda  -a_3(\chi)=0,
\label{eq:poly}
\eea
with the eigenvalues $\lambda_{1,2,3}$ sorted from the smallest in magnitude to largest by their real part. 
The coefficients of the polynomial are given by  
%
\bea
a_1&=&w_{0,0}+w_{1,1}+w_{2,2},
\nonumber\\
a_2(\chi)&=&
w_{0,0}w_{1,1} + w_{0,0}w_{2,2} + w_{1,1}w_{2,2}
\nonumber\\
&-& w_{0,1}(\chi_c)w_{1,0}(\chi_c)- w_{0,2}w_{2,0}
-w_{1,2}(\chi_w)w_{2,1}(\chi_w),
\nonumber\\
a_3(\chi)&=&  w_{0,0}w_{1,1} w_{2,2} -   w_{0,0}w_{1,2}(\chi_w) w_{2,1}(\chi_w)
\nonumber\\
&-&  w_{0,1}(\chi_c)w_{1,0}(\chi_c) w_{2,2}+ w_{0,1}(\chi_c)w_{1,2}(\chi_w) w_{2,0}
\nonumber\\
&+& w_{0,2}w_{2,1}(\chi_w) w_{1,0}(\chi_c)     -  w_{0,2}w_{2,0} w_{1,1}.
\eea
Here, $w_{m,n}$ are the matrix elements of $\hat W(\chi)$ in Eq. (\ref{eq:M}), with the rows and columns counted
by 0,1,2.
It is important to note that the counting field dependence disappears  within the products $w_{0,1}(\chi_c)w_{1,0}(\chi_c)$
and $w_{1,2}(\chi_w) w_{2,1}(\chi_w)$.
As a result, $a_2$ in fact does not depend on the counting fields.
Explicitly, in terms of the transition rate constants we find that
\bea
a_1&=&
-\Gamma_c\left(2n_c+1\right) -\Gamma_h\left(2n_h+1\right)- \Gamma_w\left(2n_w+1\right),
\nonumber\\
a_2&=&
\left( \Gamma_cn_c+\Gamma_hn_h \right)  \left[\Gamma_c(n_c+1) +\Gamma_wn_w \right] 
\nonumber\\
&+&\left[\Gamma_cn_c+\Gamma_hn_h\right]   \left[\Gamma_w(n_w+1) +\Gamma_h(n_h+1)\right] 
\nonumber\\
&+&\left[\Gamma_c(n_c+1)+\Gamma_wn_w\right] \left[\Gamma_w(n_w+1) +\Gamma_h(n_h+1)\right]  
\nonumber\\
&-&\Gamma_c^2(n_c+1)n_c -\Gamma_h^2(n_h+1)n_h  
 -\Gamma_w^2(n_w+1)n_w. 
\nonumber\\
\label{eq:a1a2}
\eea
%
%
The coefficient $a_3$ has two contributions that depend on $\chi$,
 while the rest of the expression is lumped into the constant $C$,
\bea
a_3(\chi)&=&w_{0,1}(\chi_c)w_{1,2}(\chi_w) w_{2,0} +  w_{0,2}w_{2,1}(\chi_w) w_{1,0}(\chi_c)  + C
\nonumber\\
&=&\Gamma_c\Gamma_w\Gamma_h(n_c+1)e^{-i\theta_c\chi_c} (n_w+1) e^{-i\theta_w\chi_w}n_h
\nonumber\\
&+&\Gamma_c\Gamma_w\Gamma_h(n_h +1)  n_wn_ce^{i\theta_c\chi_c}  e^{i\theta_w\chi_w} + C.
\label{eq:a3}
\eea
It is useful to note that  $a_2>0$ and that $a_3(\chi=0)=0$.

The cumulant generating function $G(\chi)$ is  formally defined as
\bea G(\chi)= \lim_{t \to \infty} \ \frac{1}{t}\ln\sum_{j}\sum_{n,m} \mathcal
P_t(j,n\theta_c,m\theta_w)e^{in\theta_c\chi_c}e^{im\theta_w\chi_w}. \label{eq:Gd2}
\nonumber\\
\eea
%
In terms of the characteristic function it is given by
\bea
G(\chi) =  \lim_{t \to \infty} \ \frac{1}{t}\ln \langle I| Z(\chi,t) \rangle,
\eea
with $\langle I|= \langle 1 1 1|$ a unit left vector.
To obtain the CGF  we diagonalize $\hat W(\chi)$ and extract the eigenvalue that 
dictates the long-time dynamics---this is the eigenvalue with the smallest magnitude for its real part,
\bea
G(\chi)=\lambda_1(\chi).
\label{eq:lam1}
\eea
We refer to this solution as the ``complete CGF". 
For  3$\times 3$ or larger matrices the diagonalization of $\hat W$ is performed numerically, 
possibly suffering from accuracy and stability issues.
It is easy to verify that $a_3$ satisfies the exchange 
fluctuation symmetry
\bea
a_3(\chi_c,\chi_w)= a_3(i(\beta_h-\beta_c)-\chi_c, i(\beta_h-\beta_w)-\chi_w),
\nonumber\\
\label{eq:sym}
\eea
which is a microscopic verification of the second law of thermodynamics \cite{esposito-review}.
%
The CGF provides e.g. the currents ($\nu=c,w$) and the noise terms as
\bea
\langle J_{\nu} \rangle&=& \frac{\partial G}{\partial (i\chi_{\nu})}\Big|_{\chi=0},
\nonumber\\
\langle S_{\nu}\rangle &=&  \frac{\partial^2 G}{\partial (i\chi_{\nu})^2}\Big|_{\chi=0}.
\eea
While we can reach these coefficients numerically, we are interested here in closed-form expressions. 
Nevertheless,  acquiring analytically the roots of a cubic (or higher order)
characteristic function is obviously not a trivial task.
To bypass this challenge, next we illustrate that by truncating the characteristic 
equation (\ref{eq:poly}) we reach partial CGFs, which are simple to solve and
yield analytical expressions for the current and its noise.

In fact, this idea was introduced earlier on for the calculation of the diffusion coefficient in Ref. \cite{Koza}.
A recent application of this principle
in the context of chemical kinetics was discussed in Ref. \cite{Udo-Fano}.
Nevertheless, here we emphasize that partial CGFs are thermodynamically consistent (maintaining the fluctuation relation), 
show that a partial $n$th-order CGFs can approximate cumulants beyond that order, 
and derive analytical expressions for the current and its noise for the canonical three-level QAR model.

\subsection{First order CGF}
To obtain the first cumulant, namely energy currents, it is sufficient to keep only the linear term in $\lambda$
within the polynomial (\ref{eq:poly}). 
This linear order CGF satisfies the symmetry relation (\ref{eq:sym}), and it is given by
\bea
G^{(1)}(\chi_c,\chi_w)=
 \frac{a_3(\chi_c,\chi_w)}{a_2}.
 \label{eq:G1}
\eea
The heat current from the cold bath towards the system immediately follows by taking a derivative with respect
to $(i\chi_c)$ and using Eq. (\ref{eq:a3}),
\bea
\langle J_c\rangle
=-\frac{\theta_c \Gamma_c\Gamma_w\Gamma_h
\left[
(n_c+1)(n_w+1)n_h -
(n_h +1)  n_wn_c
\right]}{a_2}.
\nonumber\\
\label{eq:Jc}
\eea
We proceed and derive a cooling condition, $\langle J_c\rangle\geq0$,
\bea
e^{-\beta_c\theta_c} e^{-\beta_w\theta_w} - e^{-\beta_h\theta_h} \geq 0,
\eea
which translates to
\bea
\frac{\theta_c}{\theta_h} \leq \frac{\beta_h-\beta_w}{\beta_c-\beta_w}.
\label{eq:cond}
\eea
We can similarly organize an expression for 
$\langle J_w\rangle $: Back to  Eq. (\ref{eq:G1}), we take the first derivative with respect to $(i\chi_w)$.
It is obvious that $\langle J_w\rangle $
is given by (\ref{eq:Jc})---with $\theta_w$ replacing $\theta_c$.
As a result, the coefficient of performance (sometimes termed cooling efficiency) reduces to
\bea
\eta \equiv \langle J_c\rangle/\langle J_w\rangle= \theta_c/\theta_w.
\eea
While these results were received before \cite{reviewARPC14,Levy12,joseSR},
in our approach we accomplish them rather instantly.
It is important to note that since we truncate the characteristic function to linear order in $\lambda$,
our results are accurate for the heat current, 
but the current noise is incomplete.

Reorganizing the cooling condition (\ref{eq:cond}) we get
\bea
\theta_c/\theta_w\leq(\beta_{h}-\beta_w)/(\beta_c-\beta_h), 
\eea
which translates
to $\eta\leq \eta_c$, with the Carnot COP identified as
$\eta_c=\frac{\beta_h-\beta_w}{\beta_c-\beta_h}$. 
This bound was derived in Ref. \cite{bijayS} based on a full counting statistics approach.
This maximal COP is achieved when the cooling current (\ref{eq:Jc}) diminishes. 
The heat currents  $\langle J_{h,w}\rangle$ nullify at that point as well, with zero entropy production rate.
Beyond that, the three-level QAR described in this work is an endoreversible heat machine:
finite heat transfer rates 
between the working fluid and the reservoirs reduce the efficiency below the Carnot bound.

\begin{figure*}[tbp]
\hspace{-20mm}
\includegraphics[width=16cm]{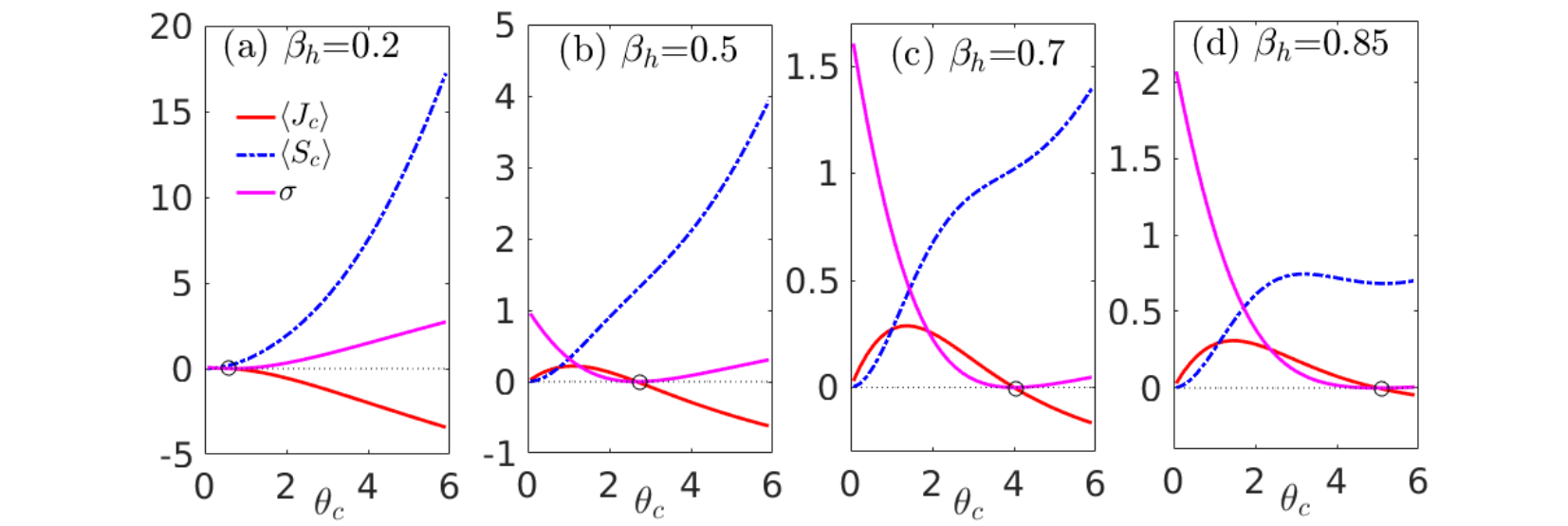}
\caption{(a)-(d) Cooling current and noise of the three-level QAR as a function of $\theta_c=\theta_h-\theta_w$.
We fix $\beta_c=1$, $\beta_w=0.1$, $\theta_h=6$, and tune $\beta_h$ from high (panel a) to low (panel (d))  temperature.
The boundary of the cooling window is marked by a circle.
}
\label{FigJS}
\end{figure*}

\begin{figure*}[tbp]
\hspace{-20mm}
\includegraphics[width=16cm]{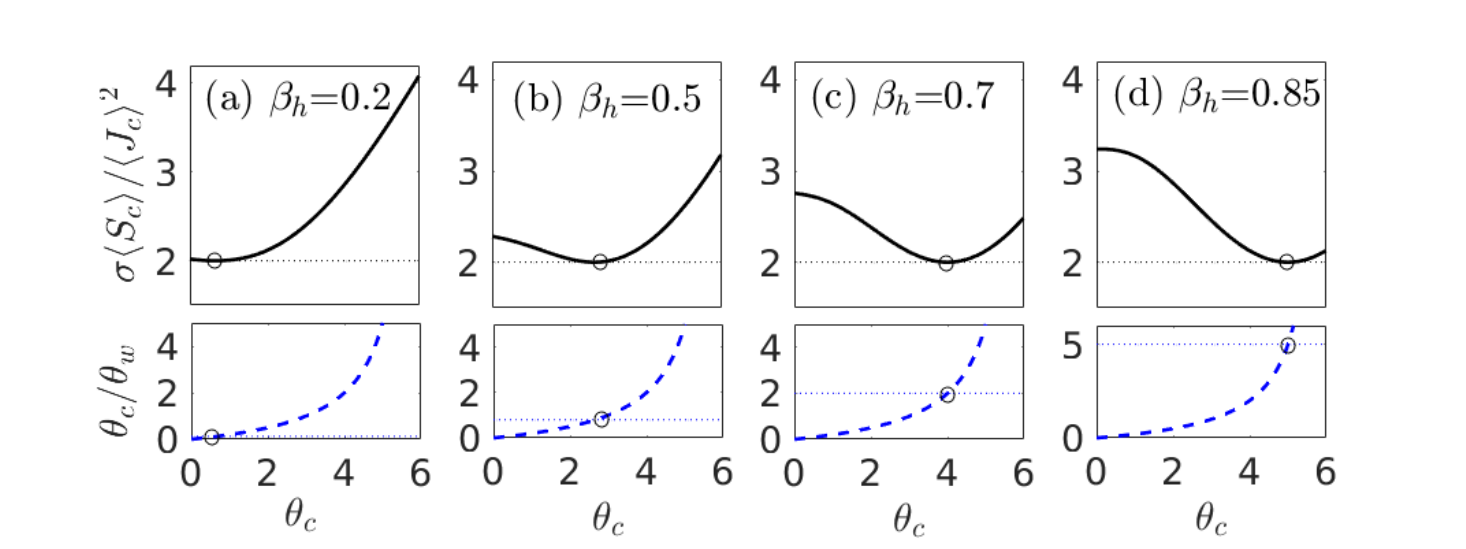}  
\caption{(a)-(d) Thermodynamic uncertainty relation (full) along with its bound of 2 (dotted),
$k_B\equiv1$.
In the bottom panels we display the coefficient of performance $\theta_c/\theta_w$ (dashed) and its maximal Carnot 
value (dotted),
$\eta_C=(\beta_h-\beta_w)/(\beta_c-\beta_h)$.
Parameters are the same as in Fig. \ref{FigJS}.
The boundary of the cooling window is marked by a circle.
}
\label{FigTUR1}
\end{figure*}

\subsection{Second order CGF}
To derive a closed-form expression for the second cumulant, the current noise,
we need to retain the quadratic term in the polynomial equation
(\ref{eq:poly}),
\bea
\lambda^2 - \frac{a_2}{a_1}\lambda  +\frac{a_3(\chi)}{a_1}=0.
\label{eq:poly2}
\eea
%
%
The resulting-partial CGF is given by the root with the smallest-magnitude real part,
%
\bea G^{(2)}(\chi)=\frac{a_2 - \sqrt{a_2^2-4a_3(\chi)a_1}}{2a_1}.
\label{eq:G2}
 \eea
This partial CGF preserves the fluctuation symmetry.
By taking the first and second derivatives with respect to $\chi_c$, we derive expressions
for the cooling current and its noise,
\bea
\langle J_c \rangle&=& \frac{1}{a_2} \frac{\partial a_3}{\partial (i\chi_c)}\Big|_{\chi=0},
\nonumber\\
\langle S_c \rangle &=& \frac{1}{a_2} \left[ \frac{\partial^2 a_3
}{\partial (i\chi_c)^2} + \frac{2a_1}{a_2^2} \left( \frac{\partial
a_3}{\partial (i\chi_c)}\right)^2\right]\Bigg|_{\chi=0},
\label{eq:JS2}
 \eea
where
\bea
\frac{\partial a_3}{\partial (i\chi_c)} \Big|_{\chi=0}&=&
-\theta_c
\Gamma_c \Gamma_h \Gamma_w
\\
&\times&
\left[ (n_c+1)(n_w+1) n_h
-(n_h +1)  n_wn_c \right],
\nonumber
\eea
and
\bea
\frac{\partial^2 a_3}{\partial (i\chi_c)^2}\Big|_{\chi=0} &=&
\theta_c^2 \Gamma_c \Gamma_h \Gamma_w
\\
&\times&\left[ (n_c+1)(n_w+1) n_h
+(n_h +1)  n_wn_c \right].
\nonumber
\eea
The cooling current agrees with  Eq. (\ref{eq:Jc}). It nullifies at
(i) thermal equilibrium or (ii) at the maximum (Carnot) cooling efficiency.
We can similarly calculate the current noise at the other contacts.

Eqs. (\ref{eq:G2})-(\ref{eq:JS2}) constitute the main results of this work. 
By truncating the characteristic polynomial to second order in $\lambda$
we retain all linear and quadratic terms in $\chi$ within (\ref{eq:poly2}).
This equation can be readily solved, to hand over  closed-form expressions for the current and its 
noise---for multilevel systems.
This simple principle can be employed in more intriguing cases, 
e.g., when quantum coherences between states persist in the steady state limit \cite{KilgourD}.
It is also worth pointing out that while a fully numerical solution of the eigenvalue problem may become 
unstable in certain parameters,
the calculation of the first two cumulants as suggested here is simple, robust---and exact.


Besides current and its noise, we calculate the total entropy production rate in the system
\bea
\sigma =\sum_{\nu=c,h,w} -\frac{\langle J_{\nu}\rangle}{T_{\nu}}.
\eea
An intriguing ``thermodynamic uncertainty relation" (TUR) was recently discovered for 
Markov processes in steady state. It
relates the averaged energy current, its fluctuations, and the entropy production rate in
the nonequilibrium process $\sigma$ \cite{TUR1,TUR2,TUR3a,TUR3b},
\bea
\frac{\langle S_{\nu}\rangle}{  \langle J_{\nu}\rangle^2} \frac{\sigma }{ k_B} \geq 2.
\label{eq:TUR}
\eea
Close to equilibrium this relation collapses to the Green-Kubo equality of linear response theory.
The bound points to a crucial trade-off between precision and dissipation:
A precise process with a little noise requires high thermodynamic-entropic cost.
Since our approach relies on a thermodynamically-consistent Markov process, the TUR is satisfied in this case.
Nevertheless, it is interesting to estimate this trade-off relation
for the three-level QAR. 



\begin{figure}[htbp]
\hspace{-3mm}
\includegraphics[width=9cm]{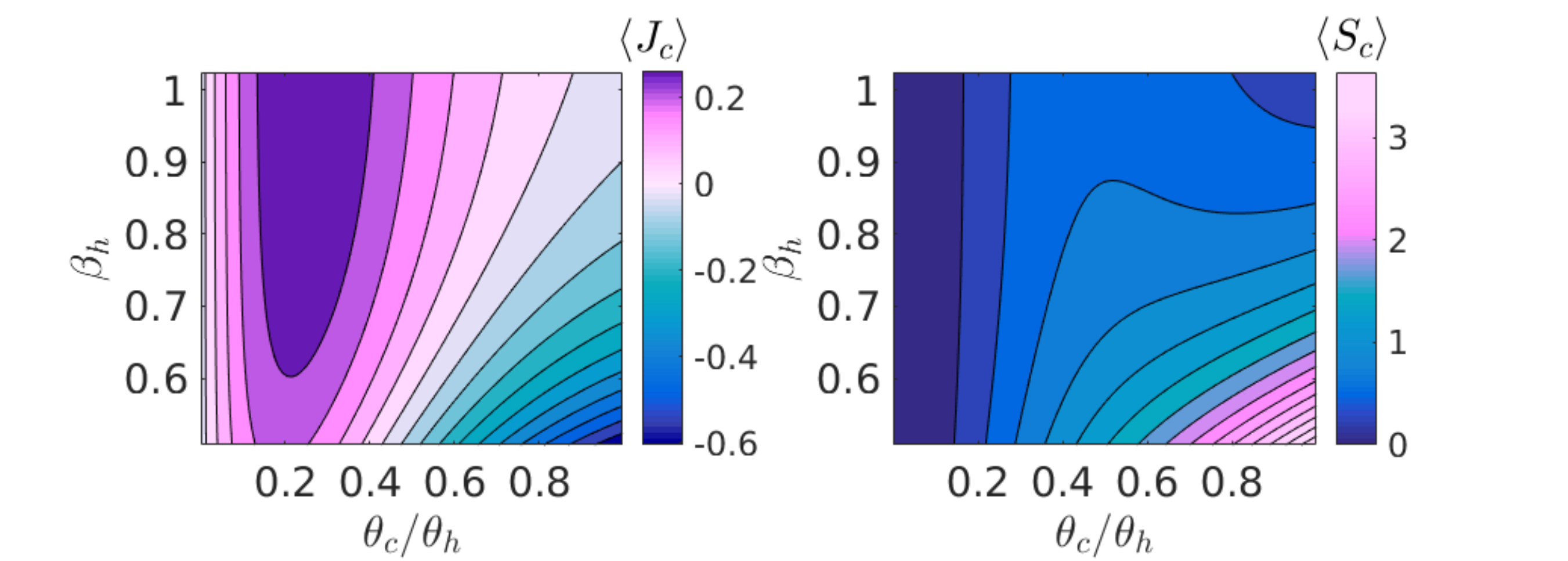} 
\caption{Contour map of (a) cooling current and (b) noise of the three-level
QAR as a function of $\theta_c$
and  $\beta_h$.
We fix $\beta_c=1$, $\beta_w=0.1$, $\theta_h=6$.
}
\label{FigC}
\end{figure}

\begin{figure*}[htbp]
\includegraphics[width=15cm]{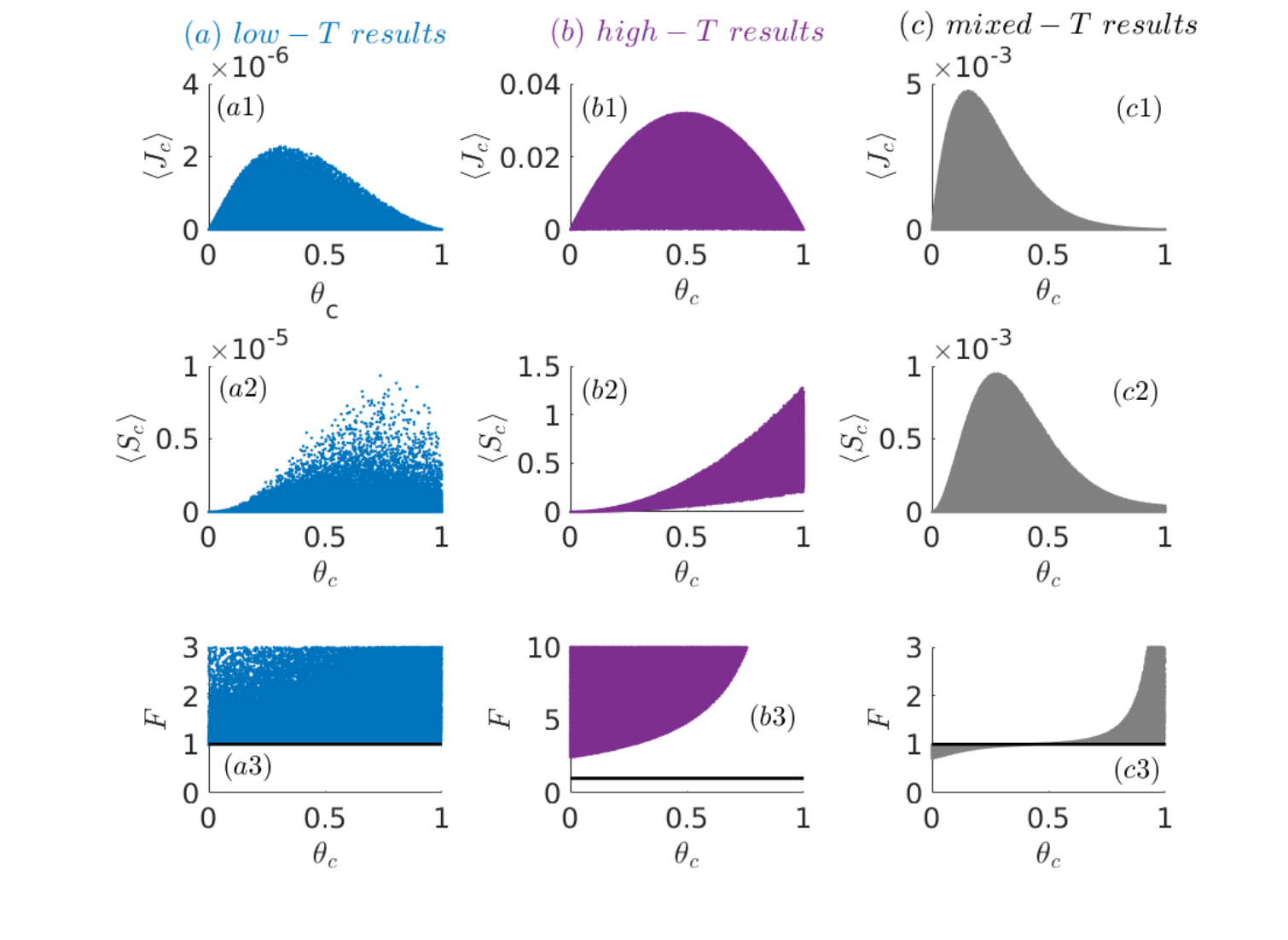}
\vspace{-7mm}
\caption{
Current, noise and the Fano factor $F\equiv \frac{\langle S_c\rangle}{\theta_c \langle J_c\rangle}$
 in the quantum, classical and mixed regimes of operation.
(a) Low-$T$  regime. Temperatures are random numbers sampled within the interval [0, 0.1], while enforcing $T_w>T_c=T_h$.
(b) High-$T$ regime. Temperatures are random  numbers within the range [1,  6], while enforcing $T_w>T_c=T_h$.
(c) Mixed-$T$ regime. We generate random numbers between [0, 0.1] for $T_c=T_h$, and random numbers
in the range [1, 6] for $T_w$.
Other parameters are $\theta_h=1$. Each panel displays $10^6$ random configurations. 
}
\label{Figrand}
\end{figure*}


\section{Simulations} 
\label{Ssimul}
What quantities characterize the performance of QARs?
Observables explored so far in the literature concern the first cumulant of the CGF, which corresponds to
the averaged current.
Specifically, designs for QARs were analyzed and optimized with respect to the cooling window, 
the averaged cooling current, and the cooling coefficient of performance,
which refers to the ratio of the extracted cooling current to the input heat 
current from the work reservoir \cite{reviewARPC14,joseSR}.

Beyond the averaged currents, our formalism allows us to explore the behavior of 
fluctuations in the operation of QARs, in the steady-state limit. 
The study of fluctuations in nonequilibrium systems has proven to be exceptionally fruitful in different disciplines
such as quantum transport \cite{fluct-revButt} and biophysics \cite{fluct-revUdo},
since fluctuations can reveal intrinsic information on the underlying dynamics. 
For example, fluctuations can assist in 
identifying the architecture of a Markovian reaction network  \cite{Fano-bio,Udo-Fano}, including
revealing  the nature of the network (unicycle or multicycle) and the number of intermediate states within
the cycle. Fluctuations may also expose the impact of many body interactions 
 and quantum phenomena, such as quantum coherences and correlations on transport behavior.
Moreover, as discussed above, the thermodynamics uncertainty relation allows to 
evaluate the trade-off between precision (little noise) and cost (entropy production). 

In our simulations below we use an ohmic function to model the coupling terms,
$\Gamma_{\nu}(\theta_{\nu})=\gamma_{\nu}\theta_{\nu}$.
For simplicity, we select identical coupling constants and set them all to unity, $\gamma_{\nu}=1$. 
We work in units of $\hbar\equiv 1$ and $k_B\equiv 1$. 

\vspace{5mm}
\subsection{Partial Cumulant Generating Functions}
We begin our discussion by inspecting the CGF of the three-level QAR.
We recall that the first-order partial CGF provides an exact result for the first cumulant (mean current),
while the second-order CGF administers exact results for the first two cumulants.  
It is interesting to test whether these partial CGFs can further 
hand over reasonable estimates for higher order cumulants.
In panel (a) of Fig. \ref{FigCGF} we plot
the complete CGF, $G(\chi_c)$, obtained numerically by solving the eigenvalue problem.
To present our results, we set $\chi_w=0$ and vary $\chi_c$ between $\pm \pi$.
We separately display the real part of the CGF, which is even in $\chi_c$,
and the imaginary part, which is odd in $\chi_c$.

We found (not shown) that the first- and second-order partial CGFs essentially look the same 
as the complete CGF to the naked eye. Therefore, to expose errors associated with partial CGFs we display
the deviation of partial CGFs from the complete function, see panels (b) and (c).
As expected, around $\chi_c=0$, or more precisely, when $\chi_c/(2\pi)\ll 1$, the partial CGFs match the exact CGF.
Beyond that, deviations of the order of
20\% are exhibited for $G^{(1)}(\chi_c)$ when $\chi_c$ is large.
The second order function $G^{(2)}(\chi_c)$ is quite accurate even for large $\chi_c$, and the error stays 
below 0.5\% throughout. 

Overall, we conclude that the second-order partial CGF not only provides an exact solution for the first two cumulants, but it
further prepares an excellent approximation to higher
order cumulants (skewness and beyond). 
It should be noted that the cooling window for the presented parameters satisfies  
$\theta_c/\theta_h\leq 0.625$. Interestingly,
the deviation of the partial CGF from the exact one, as displayed in panels (b)-(c), 
is most significant inside the cooling window around $\theta_c/\theta_h=0.2$. 

\subsection{Performance Measures}
The performance of the QAR can be optimized with respect to different observables: 
current, cooling efficiency, noise. As well,
the thermodynamic uncertainty relation (\ref{eq:TUR}) connects current fluctuations with the thermodynamic cost (dissipation).

In Figs. \ref{FigJS}-\ref{FigTUR1} we exemplify the behavior of these quantities,
displayed as a function of $\theta_c$, while tuning the temperature of the 
hot bath from high, $T_w>T_h\gg T_c$, to low, $T_w\gg T_h>T_c$.
The system acts as a chiller when $\langle J_c\rangle >0$, which is the case for small enough $\theta_c$. 
It crosses into a no-cooling region outside the window of Eq. (\ref{eq:cond}). 
In Figs. \ref{FigJS}-\ref{FigTUR1} we present results inside and outside the cooling regime so as to provide a complete picture 
over the behavior of current and noise in the QAR.

We circle the reversible point at which $\langle J_{\nu}\rangle=0$,
the entropy production vanishes, the Carnot COP is achieved (dotted line
in Fig. \ref{FigTUR1}), and
the TUR approaches the bound, which is nothing but the Green-Kubo relation.
Away from the reversible point the entropy production rate is finite, the noise may be increased or reduced,
yet overall the TUR is satisfied, as expected for a Markov process.
In Fig. \ref{FigTUR1}, the dashed line depicts the cooling COP, $\eta=\theta_c/\theta_w$.
When $\theta_c/\theta_w$ exceeds $\eta_c$, the system ceases to cool,
see Fig. \ref{FigJS}. 
Specifically, when $\beta_h\to \beta_w$ (panel (a) of Figs. \ref{FigJS}-\ref{FigTUR1})
the system operates as a chiller only at very small values for $\theta_c$.
In this regime $\langle S_c\rangle$ grows monotonically. 
As we reduce $T_h$ towards $T_c$, panels (b)-(d) display a nontrivial behavior for the current noise:
upon approaching the maximal efficiency,
when the cooling current diminishes to zero, 
the noise may be reduced as well, see panel (d). 
Altogether, Figs. \ref{FigJS}-\ref{FigTUR1} present key quantities that should be considered for the optimization
of cooling devices: the cooling current, its noise, dissipation,
and the coefficient of performance.

The non-monotonic behavior of the current and its noise are further exemplified in Fig. \ref{FigC},
where we scan over $\theta_c$ and $\beta_h$.
We verified numerically that these two cumulants, which were obtained from Eq. (\ref{eq:JS2}),
exactly match the solution of the complete CGF, where we receive the cumulants by performing 
numerical derivatives.

\subsection{Limits and Bounds}

So far, we examined the behavior of the cooling current and its noise in particular cases.
Moving beyond examples, we would like to understand the behavior of the current and its noise in typical operational limits.
Some of the questions that we ask are:
What is the maximal cooling current, or the minimal noise that may be observed in the three-level QAR?
Are there bounds on the noise or the Fano factor, defined here as a dimensionless parameter,
$F\equiv \frac{\langle S_c\rangle}{\theta_c \langle J_c\rangle}$? Can the current and its noise expose the quantumness of the device?
As we demonstrate next, indeed the first two cumulants reflect on the quantumness of the network---presented in this model
through the quantum statistics of the reservoirs.

The expressions for the current and noise are quite cumbersome, 
and it is difficult to analytically resolve limits and bounds for performance.
Let us instead numerically construct many different QAR configurations.
All setups take the same total gap $\theta_h$, but they differ in the
frequency $\theta_c$ and the temperatures $T_{\nu}$ while satisfying (for simplicity) $T_c=T_h <T_w$.
The current,  noise and the Fano factor of this ensemble of machines 
are presented in Fig. \ref{Figrand}. Every point corresponds to a particular setup
with its own combination of temperatures and frequency $\theta_c$, while extracting energy from the $c$ bath.
Note that for simplicity we set $\theta_h=1$. 
The parameters 
are sampled from a uniform distribution.
In panel (a), $T_{\nu}\ll \theta_h$;
in panel (b), the reservoirs' temperatures are high, $T_{\nu}>\theta_h$;
in panel (c) we study the mixed scenario with $T_w>\theta_h>T_c$, $T_c=T_h$.

Fig.  \ref{Figrand} shows rich characteristics; the current and noise display distinctive features in the three 
regimes of high-$T$, low-$T$, and mixed-$T$:
(i) The current is maximized at different values for $\theta_c$, and it is bounded by different functional forms.
(ii) Similarly, the noise behaves differently in the three cases.
(iii) $F\geq 1$
in the quantum limit, $F>1$ in the classical case, but
$F$ may receive values below unity in the mixed case.

It is intriguing to try and resolve the distinctive functional forms analytically.
Specifically, it is evident from panel (b1) that in the classical high-$T$ regime,
$\langle J_c\rangle \leq C \theta_c(\theta_h-\theta_c)$, with $C$ as a constant.
Studying the cooling current (\ref{eq:Jc}) 
at high temperatures by making the substitution
$n_{\alpha}(\theta_{\alpha})\to T_{\alpha}/\theta_{\alpha}$, we get
\begin{widetext}
\bea
\langle J_c\rangle = 
T_c\theta_c\frac{(T_w-T_c-\theta_c)(\theta_h-\theta_c)}
{ 3T_c(2T_w+T_c+\theta_w) + \theta_c( 2T_c+T_w+\theta_h+\theta_w) + \theta_h(2T_c+T_w)}.
\eea
\end{widetext}
This relation allows us to upper bound the current,
\bea
\langle J_c\rangle < 
\theta_c(\theta_h-\theta_c) \frac{(T_w-T_c-\theta_c)}{6T_w}.
\eea
Furthermore, if $(T_{w}-T_c) >\theta_{c}$, 
the function $\langle J_c\rangle \propto \theta_c(\theta_h-\theta_c) $ bounds the current,
in agreement with numerical simulations.
This function takes a maximal value when $\theta_c=\theta_h/2$
as we clearly observe in panel (b1). 
The noise  in panel (b2) is lower- and upper-bounded,
$A\leq\frac{\langle S_c \rangle}{\theta_c^2}\leq B$. Based on numerical simulations we found that
 $A\propto T_c$, while $B$ corresponds to the highest temperature, $B\propto T_w$. 

At low temperatures our numerical simulations demonstrate that the maximal current 
is achieved at $\theta_c\approx\theta_h/3$, while in the mixed case it is arrived at $\theta_c\approx 0.15\theta_h$. 
The noise $\langle S_c \rangle $  further displays distinctive features in the three different regimes.
Altogether, Fig. \ref{Figrand} exposes that the ensembles of classical, quantum and mixed chillers follow different operation bounds.

The three-level QAR examined in this work does not rely on internal quantum features. 
The setup is referred to as ``quantum" given the discreteness of the working fluid (three level system), 
and the quantum statistics employed for the reservoirs (Bose-Einstein distribution).
Moving forward, we envisage that the survival of steady-state coherences  in more complex models \cite{KilgourD} 
may  be exhibited within their noise characteristics.



\section{Summary}
\label{Ssum}

We described a full counting statistics formalism 
for the calculation of the cooling current and its noise in a
QAR model weakly coupled to three thermal reservoirs.
Achieving a closed-form analytical expression for the CGF 
is generally a formidable (or impossible) task for multilevel machines.
However, the first two cumulants can be correctly derived by
truncating the characteristic polynomial and working with a quadratic equation. 
While we assumed the dynamics to follow a Markovian, weak-coupling, secular quantum master equation, the procedure 
outlined here could be used within more sophisticated frameworks such as the nonequilibrium
polaron-transformed Redfield equation \cite{Cao1,Cao2}. 
With the recent manifestations of a single-atom heat engine \cite{singleatom} and a three-ion QAR \cite{QARE},
measurements of current fluctuations in nanoscale machines are conceivable.

Partial CGFs could be analogously constructed to
treat charge transport problems in electronic conductors and photovoltaics devices
\cite{Antti1,Belzig,Novotny,Cao14,bijay,Hava17},
and to examine performance bounds in biochemical reaction networks \cite{TUR1,Udo,Wolde}.
Our work here was focused on the zero-frequency current noise. 
Partial CGF could be similarly used to obtain
the full spectrum of the noise in multi-level systems \cite{junjie}.
%
Future work will be dedicated to the investigation of noise and efficiency bounds in
multilevel refrigerators where quantum coherences play a role \cite{KilgourD}, and to the study of performance bounds
under strong system-bath coupling effects. \\ \\

\begin{acknowledgments}
DS acknowledges support from an NSERC Discovery Grant and the Canada Research Chair program.
\end{acknowledgments}



\begin{thebibliography}{0}

\bibitem{AR19}
J. M. Gordon and K. C. Ng,
Cool Thermodynamics
Cambridge, UK: Cambridge Int. Sci. 2000.

\bibitem{review1}
S. Vinjanampathy and J. Anders,
Quantum Thermodynamics,
Contemporary Physics {\bf 57}, 1 Taylor and Francis (2016).

\bibitem{kos13}
R. Kosloff, Quantum Thermodynamics: A Dynamical Viewpoint,
Entropy {\bf 15}, 2100 (2013).

\bibitem{reviewARPC14}
R. Kosloff and A. Levy,
Quantum Heat Engines and Refrigerators: Continuous Devices,
Ann. Rev. Phys. Chem. {\bf 65}, 365 (2014). 

\bibitem{Goold}
J. Goold, M. Huber, A. Riera, L. del Rio, and P. Skrzypczyk,
The Role of Quantum Information in Thermodynamics---a Topical Review,
J. Phys. A: Math.  Theor. {\bf 49}, 143001 (2016).

\bibitem{kos18}
R. Alicki and R. Kosloff,
Introduction to Quantum Thermodynamics: History and Prospects,
arXiv:1801.08314.


\bibitem{singleatom}
J.  Ro{\ss}nagel,  S. T. Dawkins, K. N. Tolazzi, O. Abah, E. Lutz, F. Schmidt-Kaler, K. Singer,
A Single-Atom Heat Engine,
Science  {\bf 352}, 325 (2016).

 \bibitem{squeezeE}
J. Klaers, S. Faelt, A. Imamoglu, and E. Togan,
Squeezed Thermal Reservoirs as a Resource for a Nanomechanical Engine Beyond the Carnot Limit,
Phys. Rev. X {\bf 7}, 031044 (2017).


\bibitem{QARE}
G. Maslennikov, S. Ding, R. Hablutzel, J. Gan, A. Roulet,
S. Nimmrichter, J. Dai, V. Scarani, D. Matsukevich,
Quantum Absorption Refrigerator with Trapped Ions,
 arXiv:1702.08672.

\bibitem{joseSR}
L. A. Correa, J. P. Palao, D. Alonso, and G. Adesso,
Quantum-Enhanced Absorption Refrigerators,
Sci. Rep. {\bf 4}, 3949 (2014).



\bibitem{Levy12}
A. Levy and R. Kosloff,
Quantum Absorption Refrigerator,
Phys. Rev. Lett. {\bf 108}, 070604 (2012).

\bibitem{Linden11}
P. Skrzypczyk, N. Brunner, N. Linden, and S. Popescu,
The Smallest Refrigerators Can Reach Maximal Efficiency,
J. Phys. A: Math. Theo. {\bf 44}, 492002 (2011).

\bibitem{plenio}
M. T. Mitchison, M. Huber, J. Prior, M. P. Woods, and M. B. Plenio,
Realising a Quantum Absorption Refrigerator with an Atom-Cavity System,
Quantum Science and Technology {\bf 1}, 015001 (2016).


\bibitem{PopescuPRL}
N. Linden, S. Popescu, and P. Skrzypczyk,
How Small Can Thermal Machines Be? The Smallest Possible Refrigerator,
Phys. Rev. Lett. {\bf 105}, 130401 (2010).

\bibitem{Popescu12}
N. Brunner, N. Linden, S. Popescu, and P. Skrzypczyk,
Virtual Qubits, Virtual Temperatures, and the Foundations of Thermodynamics,
Phys. Rev. E {\bf 85}, 051117 (2012).

\bibitem{Alonso14}
L. A. Correa, J. P. Palao, G. Adesso, D. Alonso,
Optimal Performance of Endoreversible Quantum Refrigerators,
Phys. Rev. E {\bf 90} (6), 062124 (2014).

\bibitem{jose15}
L. A. Correa, J. P. Palao, and D. Alonso,
Internal Dissipation and Heat Leaks in Quantum Thermodynamic Cycles,
Phys. Rev. E {\bf 92},  032136 (2015).

\bibitem{AlonsoNJP}
J. O. Gonzalez, J. P. Palao, D. Alonso,
Relation Between Topology and Heat Currents in Multilevel Absorption Machines,
N. J. Phys. {\bf 19}, 113037 (2017).




\bibitem{Anqi}
A. Mu, B. K. Agarwalla, G. Schaller, and D. Segal,
Qubit Absorption Refrigerator at Strong Coupling,
New J. Phys. {\bf 19}, 123034 (2017).


\bibitem{NitzanE1}
M. Einax and A. Nitzan,
Network Analysis of Photovoltaic Energy Conversion,
J. Phys. Chem. C {\bf 118}, 27226 (2014).




\bibitem{KilgourD}
M. Kilgour and D. Segal,
Coherence and Decoherence in Quantum Absorption Refrigerators,
arXiv:1804.10585.



\bibitem{esposito-review}
M. Esposito, U. Harbola, and S. Mukamel, 
Nonequilibrium Fluctuations, Fluctuation Theorems, and Counting Statistics in Quantum Systems,
Rev. Mod. Phys. {\bf 81}, 1665 (2009).

\bibitem{hanggi-review}
M. Campisi, P. H\"anggi, and P. Talkner, 
Colloquium: Quantum Fluctuation Relations: Foundations and Applications,
Rev. Mod. Phys. {\bf 83}, 771 (2011).

%
\bibitem{bijay-wang-review}
J.-S. Wang, B. K. Agarwalla, H. Li, and J. Thingna,
Nonequilibrium Green's Function Method for Quantum Thermal Transport,
Front. Physics {\bf 9}, 673 (2014).

\bibitem{vanden16}
K. Proesmans, B. Cleuren, C. Van den Broeck,
Power-Efficiency-Dissipation Relations in Linear Thermodynamics,
Phys. Rev. Lett. {\bf 116}, 220601 (2016).

\bibitem{TUR1}
A. C. Barato and U. Seifert,
Thermodynamic Uncertainty Relation for Biomolecular Processes,
Phys. Rev. Lett. {\bf 114}, 158101 (2015).

\bibitem{TUR2}
P. Pietzonka, A. C. Barato, and U. Seifert,
Universal Bounds on Current Fluctuations,
Phys. Rev. E {\bf 93}, 052145 (2016).

\bibitem{TUR3a}
 T. R. Gingrich, J. M.  Horowitz, N. Perunov, and J. L. England,
Dissipation Bounds All Steady State Current Fluctuations,
Phys. Rev. Lett. {\bf 116}, 120601 (2016).

\bibitem{TUR3b}
T. R. Gingrich,  G. M Rotskoff, and J. M. Horowitz,
Inferring Dissipation from Current Fluctuations,
J. Phys. A: Math. Theor. {\bf 50}, 184004 (2017).


\bibitem{Geissler}
T. R. Gingrich, G. M. Rotskoff, G. E. Crooks, and P. L. Geissler,
Near-Optimal Protocols in Complex Nonequilibrium Transformations,
Proc. Natl. Acad. Sci. USA {\bf 113}, 10263 (2016).

\bibitem{Breuer}
H.-P. Breuer and F. Petruccione,
The Theory of Open Quantum Systems,
Oxford University Press, New York 2002.


\bibitem{hava}
H. M. Friedman, B. K. Agarwalla, and D. Segal, 
Quantum Thermodynamics From Weak to Strong System-Bath Coupling,
arXiv:1802.00511.

\bibitem{Renjie}
J. Ren, P. H\"anggi, and B. Li, 
Berry-Phase-Induced Heat Pumping and Its Impact on the Fluctuation Theorem,
Phys. Rev. Lett. {\bf 104}, 170601 (2010).


\bibitem{yelenaPRB}
L. Nicolin and D. Segal,
Quantum fluctuation theorem for heat exchange in the strong coupling regime,
Phys. Rev. B {\bf 84}, 161414 (2011).

\bibitem{yelenaCGF}
L. Nicolin and D. Segal,
Non-Equilibrium Spin-Boson Model: 
Counting Statistics and the Heat Exchange Fluctuation Theorem,
J. Chem. Phys. {\bf 135}, 164106 (2011).

\bibitem{Koza}
Z. Koza,
General Technique of Calculating the Drift Velocity
and Diffusion Coefficient in Arbitrary Periodic Systems,
J. Phys. A: Math. Gen.  {\bf 32}, 7637 (1999).


\bibitem{Udo-Fano}
A. C. Barato and U. Seifert,
Universal Bound on the Fano Factor in Enzyme Kinetics,
J. Phys. Chem. B {\bf 119}, 6555 (2015).

\bibitem{bijayS}
B. K. Agarwalla, J.-H. Jiang, and D. Segal,
Quantum Efficiency Bound for Continuous Heat Engines Coupled to Noncanonical Reservoirs,
Phys. Rev. B {\bf 96}, 104304 (2017).


\bibitem{fluct-revButt}
Ya. M. Blanter and M. B\"uttiker,
Shot Noise in Mesoscopic Conductors, Phys. Reports {\bf 336}, 1 (2000).
%

\bibitem{fluct-revUdo}
U. Seifert, Stochastic Thermodynamics, Fluctuation Theorems,
and Molecular Machines. Rep. Prog. Phys. {\bf 75}, 126001 (2012).

\bibitem{Fano-bio}
J. R. Moffitt and C. Bustamante,
Extracting Signal from Noise: Kinetic Mechanisms from a Michaelis-Menten-Like Expression for
Enzymatic Fluctuations.
FEBS J. {\bf 281}, 498 (2014).


\bibitem{Cao1}
C. Wang, J. Ren, and J. Cao, 
Nonequilibrium Energy Transfer at Nanoscale: A Unified Theory from Weak to Strong Coupling,
Sci. Rep. {\bf 5} 11787 (2015).

\bibitem{Cao2}
C. Wang, J. Ren, and J. Cao,
 Unifying Quantum Heat Transfer in a Nonequilibrium Spin-Boson Model with Full Counting Statistics,
Phys. Rev. A {\bf 95}, 023610 (2017).

\bibitem{Antti1}
A. Braggio, J. Koenig, and R. Fazio, 
Full Counting Statistics in Strongly Interacting Systems: Non-Markovian Effects,
Phys. Rev. Lett. {\bf 96}, 026805 (2006)

\bibitem{Belzig}
K. Kaasbjerg and W. Belzig,
Full Counting Statistics and Shot Noise of Cotunneling in Quantum Dots and Single-Molecule Transistors,
Phys. Rev. B {\bf 91}, 235413 (2015).

\bibitem{Novotny}
T. Novotny, 
Full Counting Statistics of Electronic Transport Through Interacting Nanosystems,
J. Comput. Electron. {\bf 12}, 375 (2013).





\bibitem{Cao14}
C. Wang, J. Ren, and J. Cao, 
Optimal Tunneling Enhances the Quantum Photovoltaic Effect in Double Quantum Dots,
New J. Phys. {\bf 16}, 045019 (2014).

\bibitem{bijay}
B. K. Agarwalla, J.-H. Jiang and D. Segal,
Full Counting Statistics of Vibrationally-Assisted Electronic Conduction: 
Transport and Fluctuations of the Thermoelectric Efficiency,
Phys. Rev. B {\bf 92}, 245418 (2015).

\bibitem{Hava17}
H. Friedman, B. K. Agarwalla, D. Segal,
Effects of Vibrational Anharmonicity on Molecular 
Electronic Conduction and Thermoelectric Efficiency,
J. Chem. Phys. {\bf 146}, 092303 (2017).

\bibitem{Udo}
Stochastic Thermodynamics of Chemical Reaction Networks,
T. Schmiedl and U. Seifert, J. Chem. Phys. 126, 044101 (2007).

\bibitem{Wolde}
C. C. Govern and P. R. ten Wolde, 
Optimal Resource Allocation in Cellular Sensing Systems, 
Proc. Natl. Acad. Sci. U.S.A. {\bf 111}, 17486 (2014).

\bibitem{junjie}
J. Liu, C. -Y. Hsieh, and J. Cao, 
Frequency-Dependent Current Noise in Quantum Heat Transfer with Full Counting Statistics,
arXiv:1708:05537.


\end{thebibliography}
\end{document}